\documentclass[%
 aip,
 amsmath,amssymb,
 reprint
]{revtex4-1}

\usepackage{graphicx}
\usepackage{dcolumn}
\usepackage{bm}
\usepackage[utf8]{inputenc}
\usepackage[T1]{fontenc}
\usepackage{mathptmx}
\usepackage{siunitx}
\usepackage{subfigure}
\usepackage{amsmath}
\usepackage{longtable}
\usepackage{appendix}

\graphicspath{{figures/}}

\begin{document}

\title{Evolution of longitudinal plasma-density profiles in discharge capillaries for plasma wakefield accelerators}
\author{J.M. Garland} 
\email{matthew.james.garland@desy.de}
\author{G. Tauscher}
\author{S. Bohlen}
\author{G.J. Boyle}
\author{R. D'Arcy}
\author{L. Goldberg}
\author{K. P\~{o}der}
\author{L. Schaper}
\author{B. Schmidt}
\author{J. Osterhoff}
\affiliation{Deutsches Elektronen-Synchrotron DESY, Notkestra\ss e 85, 22607, Hamburg, Germany}
\date{\today}

\begin{abstract}
Precise characterization and tailoring of the spatial and temporal evolution of plasma density within plasma sources is critical for realizing high-quality accelerated beams in plasma wakefield accelerators. The simultaneous use of two independent diagnostic techniques allowed the temporally and spatially resolved detection of plasma density with unprecedented sensitivity and enabled the characterization of the plasma temperature at local thermodynamic equilibrium in discharge capillaries. A common-path two-color laser interferometer for obtaining the average plasma density with a sensitivity of $\num{2e15}$~cm$^{-2}$ was developed together with a plasma emission spectrometer for analyzing spectral line broadening profiles with a resolution of $\num{5e15}$~cm$^{-3}$. Both diagnostics show good agreement when applying the spectral line broadening analysis methodology of Gigosos and Carde{\~n}oso. Measured longitudinally resolved plasma density profiles exhibit a clear temporal evolution from an initial flat-top to a Gaussian-like shape in the first microseconds as material is ejected out from the capillary, deviating from the often-desired flat-top profile. For plasma with densities of 0.5~-~$\num{2.5e17}$~cm$^{-3}$, temperatures of 1~-~7~eV were indirectly measured. These measurements pave the way for highly detailed parameter tuning in plasma sources for particle accelerators and beam optics. 
\end{abstract}

\maketitle

\section{Introduction}
\label{sec:Intro}

Recent developments in the rapidly evolving area of plasma wakefield accelerator research~\cite{tajima1979laser,PhysRevLett.54.693} have demonstrated the capability to accelerate electron bunches in cm-scale plasma structures with fields up to the 100~GVm$^{-1}$ level~\cite{blumenfeld2007energy,PhysRevLett.122.084801,litos2014high,esarey2009physics}. In the blow-out~\cite{PhysRevA.44.R6189}, or bubble regime~\cite{pukhov2002laser} -- so-called because of the complete expulsion of electrons from directly behind the wakefield driver -- the field produced by the wake can be approximated by the wave-breaking field~\cite{albritton1975cold} as $E[Vm^{-1}]\approx$~96~$\sqrt{n_0[cm^{-3}]}$ where $n_0$ is the background plasma density. Hence the forces experienced by charged particle beams are governed by the local plasma density, making the control of the plasma a critical element. 

Capillary discharges~\cite{butler2002guiding,spence2003gas,karsch2007gev,leemans2014multi} are a common solution for plasma generation in a wakefield accelerator. Such devices are designed to provide specific plasma density profiles in order to generate tailored wakefields~\cite{gonsalves2011tunable}, guide laser pulses~\cite{butler2002guiding} and focus particle beams~\cite{lindstrom2018emittance}. Tailored longitudinal and radial plasma density profiles can aid in the matching of externally-injected charged particle beams, preserving transverse and longitudinal beam quality ~\cite{marsh2005beam,ariniello2019transverse,dornmair2015emittance} and allowing extended stable wakefield propagation~\cite{PhysRevLett.77.4186,geddes2004high}. Profile shaping can facilitate the realization of internal injection~\cite{bulanov1998particle,gonsalves2011tunable}, correction of dephasing~\cite{mangles2007stability} and hosing~\cite{whittum1991electron,mehrling2017mitigation} mitigation. Furthermore, the understanding and control of active plasma lenses~\cite{panofsky1950focusing,lindstrom2018emittance} -- for the strong focusing of charged particle beams -- requires precise plasma density profile and temperature knowledge. Hence, it is essential to have a well characterized and controlled plasma profile inside the plasma source. 

This paper describes the first measurement of its type in which two complementary plasma diagnostics were used to obtain spatially- and temporally-resolved plasma-density profiles as well as spatially-averaged, temporally-resolved temperature information. A spectrometer for measuring spatially resolved broadening of plasma emission spectra~\cite{griem2012spectral} predominantly caused by the Stark effect, and a common-path two-color laser interferometer~\cite{van2018density,van2019density} for measuring the line-of-sight average plasma density, were used in conjunction. By building on previous work by Gigosos and Carde{\~n}oso~\cite{gigosos1996new, gigosos2003computer} the temperature-dependent broadening of the hydrogen Balmer-alpha spectral emission line was experimentally calibrated together with the temperature-independent plasma electron density measured by the laser interferometer to yield a temperature characterization. This temperature was subsequently used to obtain highly detailed spatial and temporal electron density characterization from spectral line broadening measurements. The unprecedented sensitivity of these measurements paves the way for detailed studies of the plasma density evolution in discharge capillaries, ultimately facilitating much finer control over future applications of plasma sources in particle accelerators.

\section{Experimental setup}

FLASHForward~\cite{aschikhin2016flashforward,d2019flashforward} is a plasma wakefield experiment based at DESY Hamburg, Germany. The FLASH accelerator~\cite{tiedtke2009soft} provides up to 1.2~GeV electron bunches with micrometer emittance and per-mille energy spread to drive a wakefield in plasma. A capillary discharge is used as the plasma source. At FLASHForward a plasma characterization experiment has been developed with the aim of obtaining precise temporally- and spatially-resolved plasma profile information.

\begin{figure}[tp]
	\includegraphics[width=0.95\linewidth]{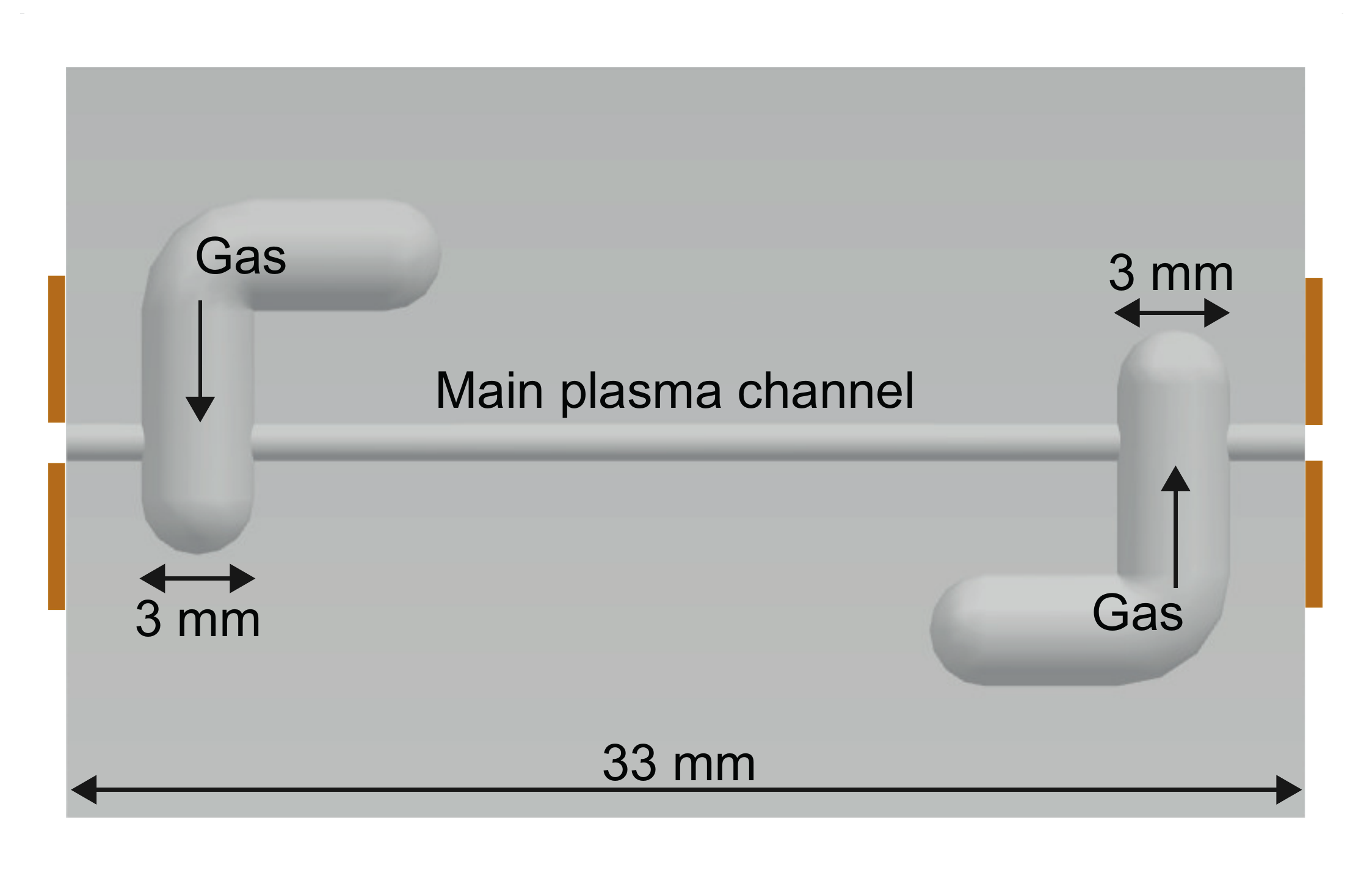}
	\caption{Schematic of the capillary discharge source used at FLASHForward. The central channel and gas inlets (opposite ends of the main channel) have a circular cross-section and are milled from sapphire. The exits of the main channel are open to allow the passage of charged particles beams and lasers. Electrodes with the same diameter opening are placed around these exits. The diameter of the main channel was either 1.0 or 1.5~mm. Spontaneous light emission was collected perpendicular to the main plasma channel axis (out of the plane of the image).}
	\label{fig:CapExp}
\end{figure}

Figure~\ref{fig:CapExp} shows a schematic of the type of capillary discharge source used at FLASHForward. When equal gas pressure is continually applied to both inlets, a constant neutral gas density profile exists in the central channel~\cite{schaper2014longitudinal}. A discharge is initiated via the two electrodes shown at the extremities of the capillary channel. In the main FLASHForward experiment the electron drive beam (used to create the wakefield) and witness beam (subsequently accelerated by the fields within the wakefield) traverse the main channel along the longitudinal axis.

As well as providing a hard-surface environment for the confinement of the plasma, the sapphire material allows the transmission of light emitted during the recombination of the plasma~\cite{gonsalves2007transverse}, hence enabling diagnostic spectroscopy. Additionally the open-ended-capillary geometry facilitates the passage of laser pulses for diagnostic purposes along the longitudinal axis. 

\begin{figure}[tp]
	\includegraphics[width=0.95\linewidth]{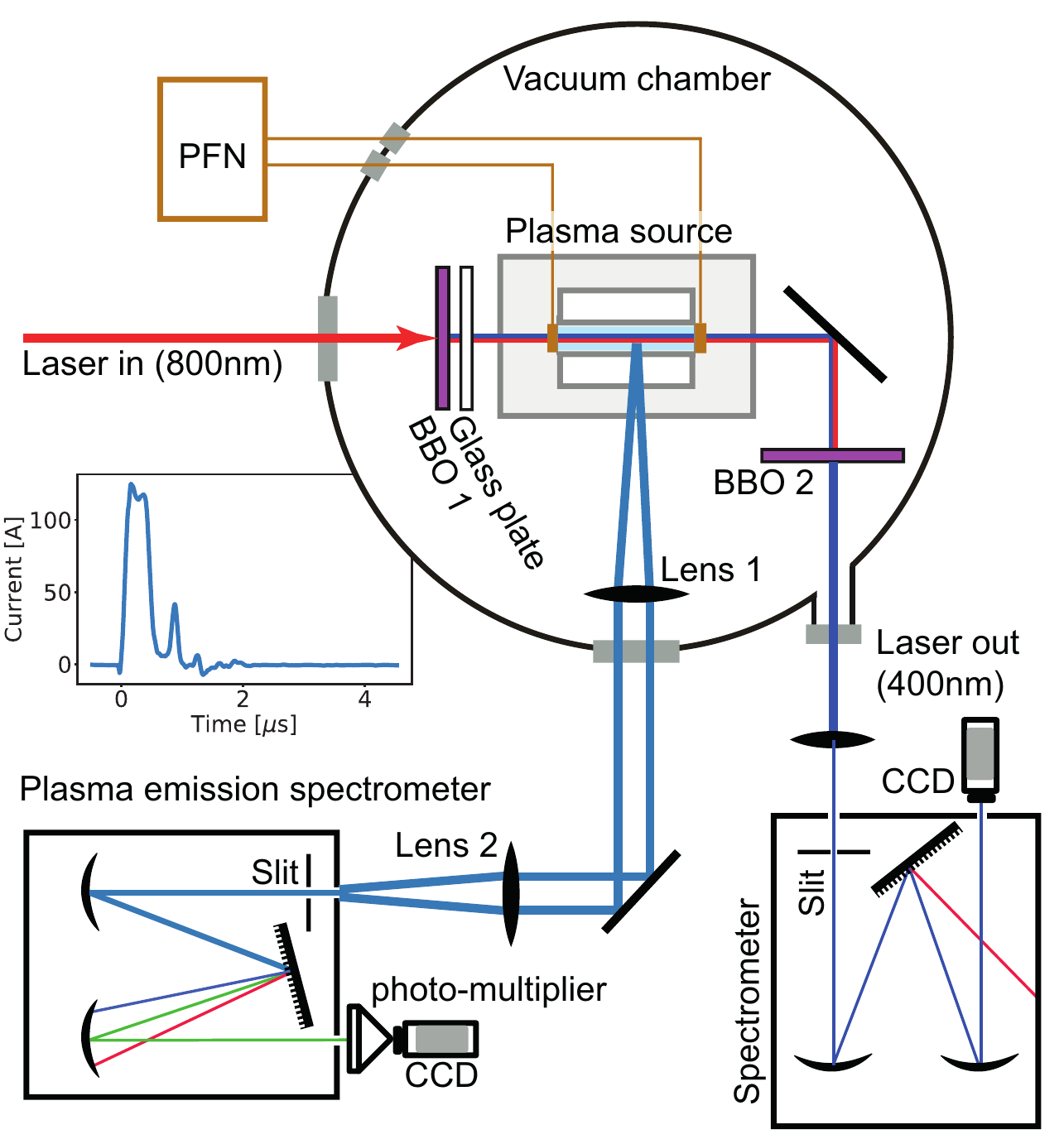}
	\caption{The experimental setup showing the two diagnostics, viewed from the top. The plasma source is mounted on a stage allowing movement along the axis of the laser. Spontaneous light emission from the plasma is collected perpendicular to the axis of the capillary channel along which the laser propagates. }
	\label{fig:expSetUp}
\end{figure}

A schematic of the experimental plasma characterization setup is shown in Fig.~\ref{fig:expSetUp}. The discharge capillary is mounted within vacuum ($\num{1e-6}$~mbar) and a pulse forming network (PFN) delivers up to 500~A of current in an almost flat-top pulse of 400~ns. The circuit matching impedance was 50$\Omega$. The voltage used to break down the gas is typically set at 25~kV. Gas is fed into the target via a buffer volume, at which location the pressure is measured using a capacitive gauge. Two diagnostics have been developed for plasma characterization in this setup and are described in the following sections.

\section{Two-color laser interferometer}
\label{sec:TCI_diag}

A common-path two-color laser interferometer (TCI) was deployed to measure the line-of-sight averaged electron density along the center of the capillary channel (see Fig.~\ref{fig:expSetUp}). The fundamental input laser wavelength is 800~nm (Titanium-sapphire). The first BBO (beta barium borate) crystal converts around 10\% of the incoming laser pulse into a second harmonic copy at 400~nm.  A 1~mm glass plate is inserted after the first BBO crystal to generate an initial temporal offset between the pulses ($\sim$150~fs). The pulses build up a relative shift during propagation through the plasma due to their different phase and group velocities. After the plasma, another fraction of the 800~nm pulse is doubled to 400~nm in the second BBO. The two perturbed pulses at the exit of the capillary are imaged onto a 10~$\mu$m slit and into a spectrometer (grating 1800~lines/mm blazed at 400~nm) after which a spectral interference pattern is observed on a CCD camera. 

\begin{figure}[tp]
\centering
	\includegraphics[width=0.93\columnwidth]{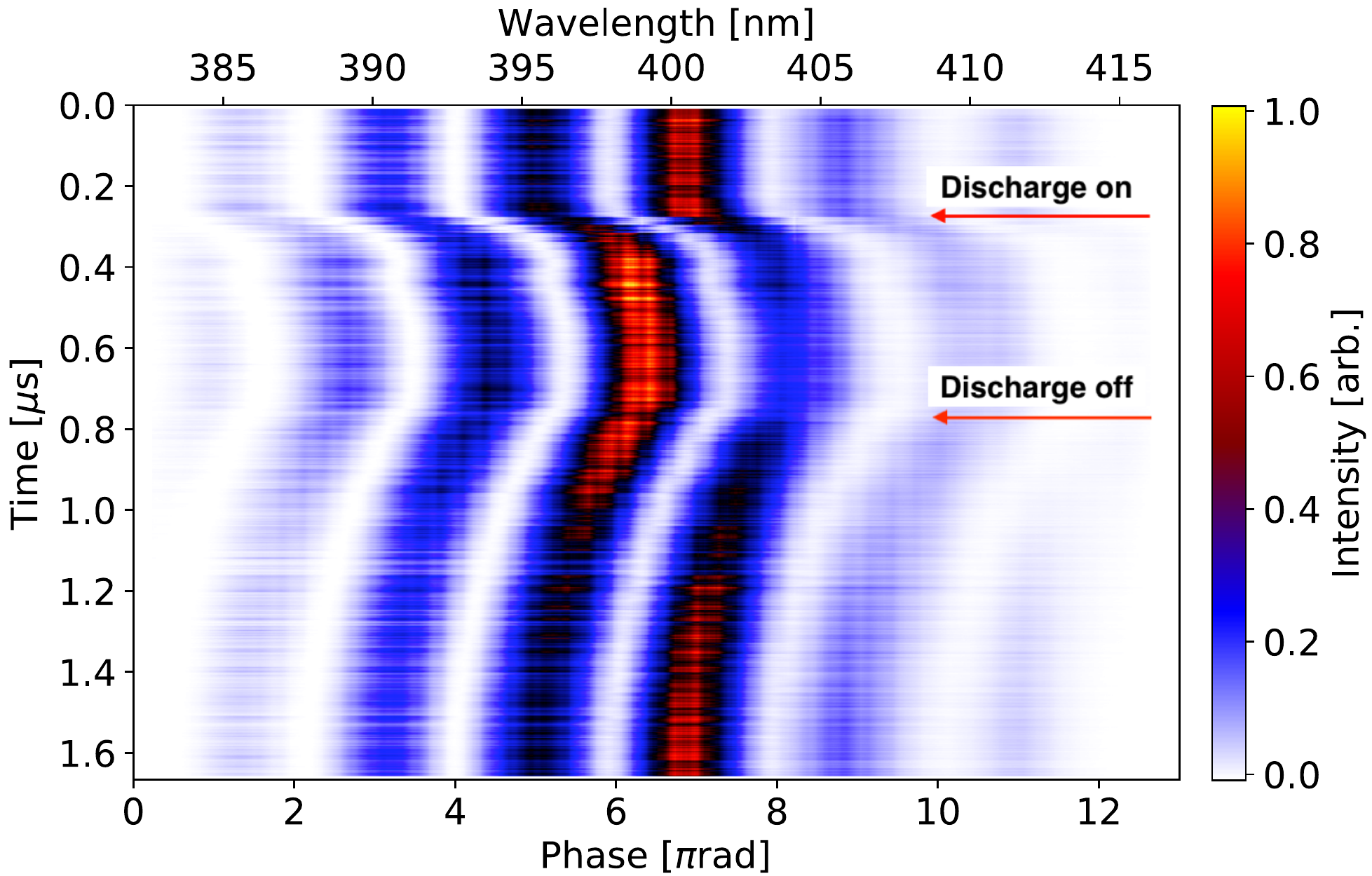}
\caption{Evolution of the interference pattern as a function of time during a discharge. Each individual interferogram is projected to the wavelength/phase axis to form one horizontal line in the waterfall plot.}
\label{fig:TCI_intPat+waterfall}
\end{figure}

Figure~\ref{fig:TCI_intPat+waterfall} shows the phase-axis projection of a set of interferograms in the form of a waterfall plot, indicating the development of the intensity maxima as a function of time as the plasma density evolves before, during, and after the discharge current pulse. The laser spot size was around 0.5~mm in the plasma and the central $\sim$0.1~mm of the spatial projection of the interferogram was selected for further analysis, corresponding to around 10\% of the capillary channel diameter. As the laser was aligned through the center of the capillary, the radial plasma density profile was therefore averaged over 0.1~mm, $\pm$0.05~mm from the longitudinal axis. 

\begin{figure}[tp]
\centering
\includegraphics[width=1\columnwidth]{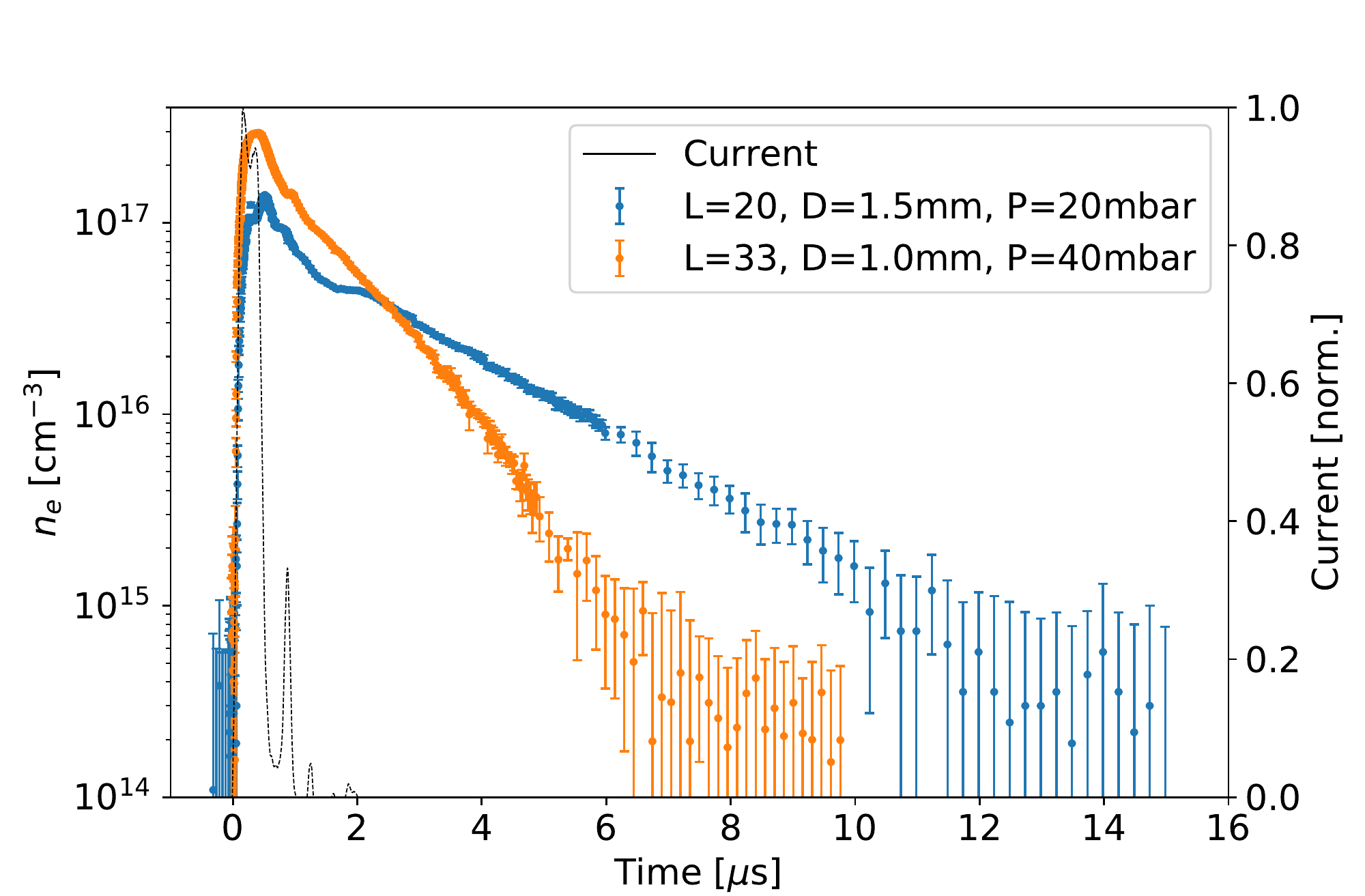}
\caption{Measurements using the TCI for capillaries of different lengths (L) and diameters (D), filled with an argon-hydrogen gas mixture at the indicated backing pressures (P). The discharge current profile is indicated by the black curve.}
\label{fig:TCI_result1}
\end{figure}

The phase shift and spacing of the interference fringes shown in Fig.~\ref{fig:TCI_intPat+waterfall}, are dependent on the phase and group velocity of the laser pulse respectively~\cite{van2018density,van2019density}. Although the spacing of the interference fringes was calculated in this work, the contribution was found to be negligible in the density range up to $\num{3e17}$~cm$^{-3}$. Therefore, the dominant phase shift contribution was used to probe the average on-axis plasma density $n_e$ by calculating the integrated phase shift as a function of time:
\begin{equation}
\label{eq:laserDen}	
	n_e = \frac{4 \epsilon_0 m_e c}{3 q_e^2} \frac{\omega_0}{L}\Delta \phi
\end{equation}
where $\omega_0$ is the fundamental wavelength of the laser (800~nm), $L$ is the length of the plasma and $\Delta\phi$ is the total integrated phase shift accumulated during propagation through the plasma~\cite{van2018density,van2019density}. The electron mass, charge, speed of light and vacuum permittivity are given by $m_e$, $q_e$, $c$ and $\epsilon_0$ respectively. 

Figure~\ref{fig:TCI_result1} shows typical measurements of the temporal evolution of the average on-axis plasma density for two different capillaries. The gas used was an argon-hydrogen mix of 95\% and 5\% by density respectively, at a steady-state gas pressure of 20 and 40~mbar respectively, measured in the buffer volume. The temporal resolution was 10~ns, limited mainly by the jitter between the laser and discharge, where the discharge jitter was the dominant factor. The plasma density resolution was limited by the instrument function of the slit, spectrometer and camera setup to $\num{2.0e14}$~cm$^{-3}$. However, as this technique relies on the integrated effect of the plasma on the laser over the plasma length $L$, the plasma density sensitivity is a function of $L$. The sensitivity can be defined as
\begin{equation}
\label{eq:laserSens}	
	\Delta n_e^{min} = \Phi/L
\end{equation}
where $\Delta n_e^{min}$ is the minimum measurable density and $\Phi$ is the sensitivity in cm$^{-2}$. The value of $\Delta n_e^{min}$ for the two capillary lengths shown in Fig.~\ref{fig:TCI_result1} can be estimated by the minimum $n_e$ measured, taking into account the upper error bound. This yields $\Delta n_e^{min} =$~$\num{1.0e15}$~cm$^{-3}$ and $\Delta n_e^{min} =$~$\num{0.6e15}$~cm$^{-3}$ for the 20 and 33~mm long capillaries respectively, which leads to a sensitivity of $\Phi =$~$\num{2.0e15}$~cm$^{-2}$.

To calculate the density in Fig.~\ref{fig:TCI_result1} the average of the total phase shift over the fixed length of the capillary channel was taken. This neglects the influence of any material ejected from the ends of the capillary along the laser path, however the diagnostic explicitly records the phase shift contribution from such expulsion. It is assumed that the material ejected in this plume spreads out rapidly in all directions into the surrounding ambient vacuum and the contribution to the line-of-sight integral outside the capillary can be considered negligible. This is supported by the density diagnostic comparison presented in Sec.~\ref{sec:DiagComp} of this work.

\section{Plasma emission spectrometer}
\label{sec:Spec_diag}

An plasma emission spectroscopy imaging diagnostic was developed (see Fig.~\ref{fig:expSetUp}) utilizing spectral line broadening~\cite{griem2012spectral} (SLB) to measure the spatially resolved electron density. The spectrometer was a Princeton Instruments SpectraPro 2150i containing a grating with 1200~lines/mm blazed at 500~nm with a spectral range of 50~nm. An Andor iStar DH334T camera was used to capture images. The camera contains an intensified photo-multiplier CCD (iCCD) system with a fast gating time down to 2~ns and a 1024~$\times$~1024 pixel array with a pixel size of 13~$\mu$m. This enabled a good temporal resolution, which was further limited by the signal/noise ratio to around 20~ns in the studies presented here. The spectrometer has a spectral resolution of 0.05~nm/pixel. The optical imaging system and CCD give a spatial resolution of around 20~$\mu$m measured with a micrometer resolution target. However, the overall resolution of the system realized during measurements is further reduced, mainly due to errors and noise present in the measurement of the spectral broadening, which is discussed later in Sec.~\ref{sec:longEvolv}. The transfer function of the spectrometer setup produces an intrinsic line broadening of 0.7$\pm$0.1~nm.

In the plasma-density range $\num{1e15}$~-~$\num{1e18}$~cm$^{-3}$ the $H_{\alpha}$ emission line in the Balmer series exhibits a temperature dependent power-law relationship between SLB and plasma electron density. Hence, small amounts of hydrogen can be added to any gas to provide tracer atoms for density diagnostic purposes. When investigating pure hydrogen, the emission spectrum is typically dominated by the $H_{\alpha}$ line within the wavelength range 630~-~680~nm. However, when analyzing spectra from other gas species doped with hydrogen great care must be taken to separate the $H_{\alpha}$ emission line from other spectral lines, which are produced by the primary gas.

\begin{figure}[tp]
\centering
\subfigure[]{
	\includegraphics[width=1\columnwidth]{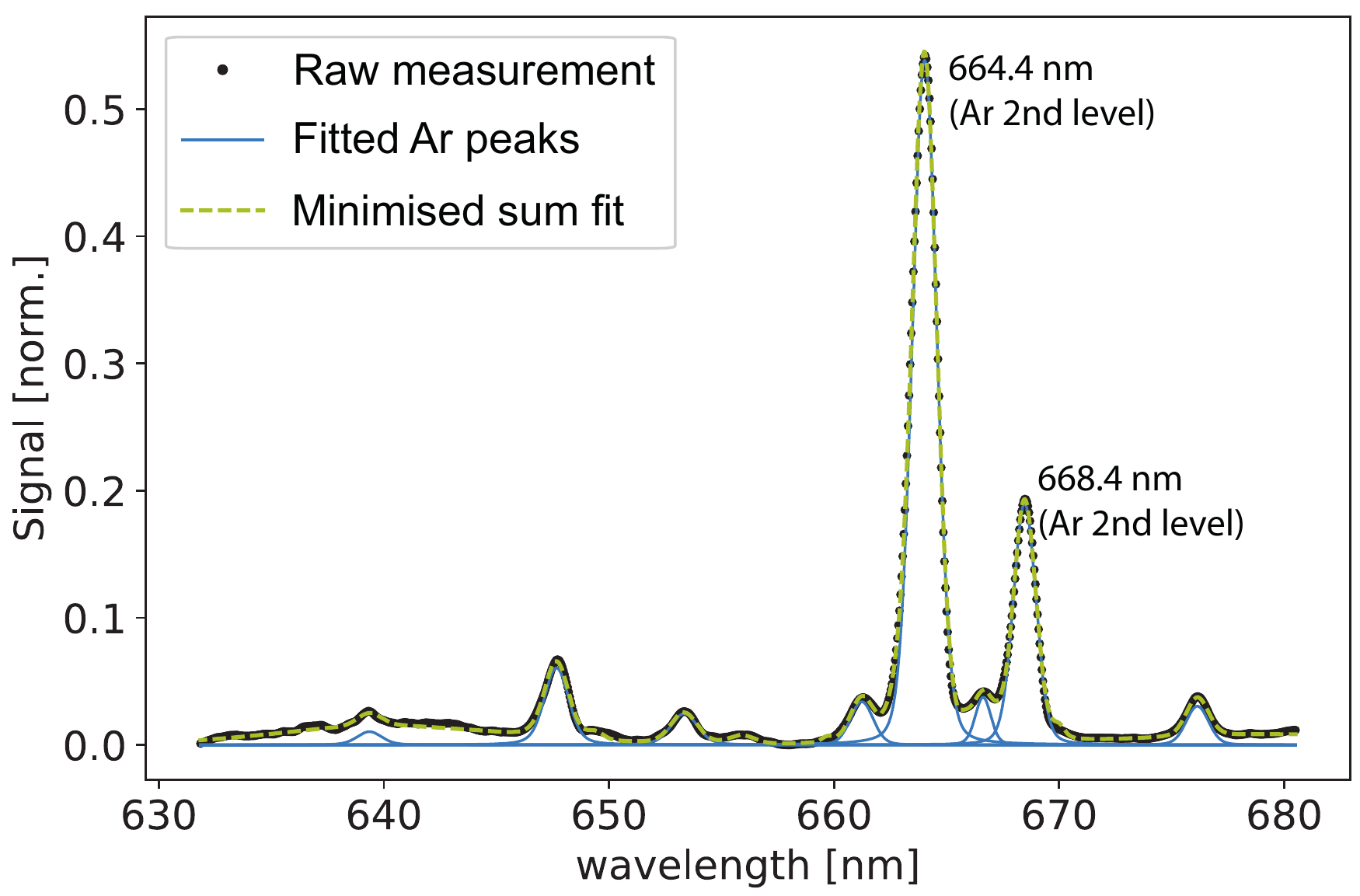}
	}
\subfigure[]{
	\includegraphics[width=1\columnwidth]{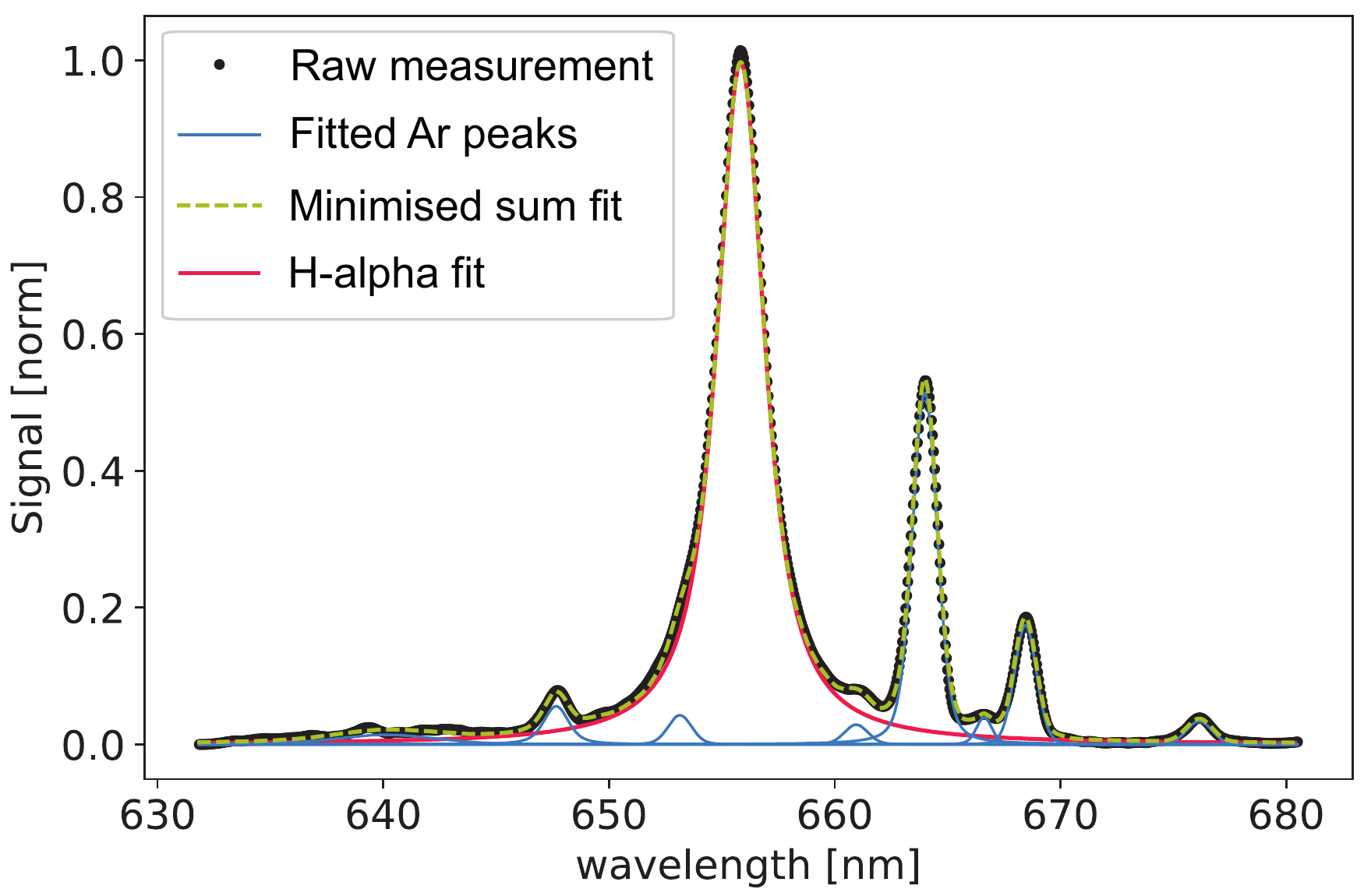}
	}	
\caption{Example spectra from pure argon (a) and argon doped with 5\% hydrogen (b). Measurement data points are shown in black and individually fitted background argon peaks in blue. The dashed yellow curve in each represents the sum of all fitted peaks with minimized deviation from the measured data. The $H_{\alpha}$ peak is shown in red in the center of (b) at a wavelength of 656.28~nm. The two larger peaks at 664.37 and 668.43 are due to the second level ionization of argon.}
\label{fig:ArSpec1}
\end{figure}

In order to identify the background emission lines as the plasma density evolves, a spectrum was first recorded in the plasma formed from pure argon, the primary gas, in the region of the $H_{\alpha}$ emission line (630~-~680~nm). Its emission lines were identified as indicated in Fig.~\ref{fig:ArSpec1}~(a). This allowed a map to be built up, which could then be used to identify and fit the lines when hydrogen was added. Figure~\ref{fig:ArSpec1}~(b) shows an example of the emission spectrum recorded in an argon-hydrogen mix of 95\% and 5\% respectively, at the same neutral gas pressure and time after the discharge as in (a). 

In the recorded spectra, the background signal is first removed by fitting the spectrum without plasma emission and subtracting this from the recorded signal with plasma emission. Additionally, the small contribution from the continuum spectrum of the plasma emission is removed by fitting a Planck function where the temperature is a free parameter. The continuum is most visible in Fig.~\ref{fig:ArSpec1}~(a) as a low-amplitude (0.02 normalized) broad signal in the wavelength range 630~-~650~nm. Subsequently, all visible peaks above 0.025 of the normalized intensity were fitted with the Faddeeva function~\cite{wells1999rapid}, which is a convolution of Gaussian and Lorentzian functions. In each fit, the minimum Gaussian component of the function was set to the instrument function resolution of 0.7~nm. A negligible additional quantity of Gaussian component was found in the density range investigated in this work, hence the Doppler broadening component due to a finite temperature did not play a significant role. Other broadening components such as fine structure, van der Waals and self-absorption were as expected, not detectable in this work. Self-absorption was negligible due to the relatively low density of the plasma ($< \num{1e18}$~cm$^{-3}$) and small line-of-sight distance (0.75~mm) through the plasma which the photons travel before collection. A sum-function of all fitted peaks was then calculated and a least-squares minimization was used to obtain the best overall sum-fit to the spectrum data (dashed yellow curves in Fig.~\ref{fig:ArSpec1}). The full width at half maximum (FWHM) of the Lorentz component of the Feddeeva function $\Delta\lambda$ was then extracted from the fit parameters of the $H_{\alpha}$ peak (red curve in Fig.~\ref{fig:ArSpec1} b) and used to calculate the electron density. The broadening resolution limit is therefore reached when no Lorentz component can be found in the solution to the Feddeeva function, and the total width of the broadened profile is equal to the instrument function broadening. The limit of the fitting residual was set to be $<$10~\% which limited the maximum detectable $\Delta\lambda$ to 0.15~nm equating to an electron density of around $\num{5e15}$~cm$^{-3}$. This resolution limit explains why no Doppler broadening component could be resolved in this work. For an electron temperature range of 1~to~10~eV, the Doppler broadening component at a wavelength of 656.28~nm is 0.05~to~0.15~nm. As will be shown in Sec.~\ref{sec:DiagComp}, no temperature above 7.6~eV was measured in this work.

The imaging system used to collect the plasma light and transport it to the spectrometer is shown in Fig.~\ref{fig:expSetUp}. The center of the capillary was positioned 150~mm from the first lens (focal length 150~mm) and the second lens (also focal length 150~mm) was used to image the center of the capillary channel onto the slit of the spectrometer; the slit opening width was set to 120~$\mu$m. The center of the capillary was imaged through the sapphire by setting the second lens position half way between the two points at which each channel-wall extremity was focused. The depth of field was approximately 10~$\mu$m. The geometry of the gas inlet regions (see Fig.~\ref{fig:CapExp}) results in a relative focal plane shift from the central axis of the capillary of up to 300~$\mu$m, due to the variable sapphire thickness. Hence, the SLB measurement in the gas inlet regions samples a radius of 0~$<$~r~$<$~300~$\mu$m from the central axis of the capillary, which is important when the radial plasma profile is non-uniform. As the field of view of the imaging system was approximately 5~mm, 11 measurements along the longitudinal dimension of the capillary were made by moving the plasma source and recombining the data in post-processing.

\begin{figure}[tp]
	\includegraphics[width=0.95\linewidth]{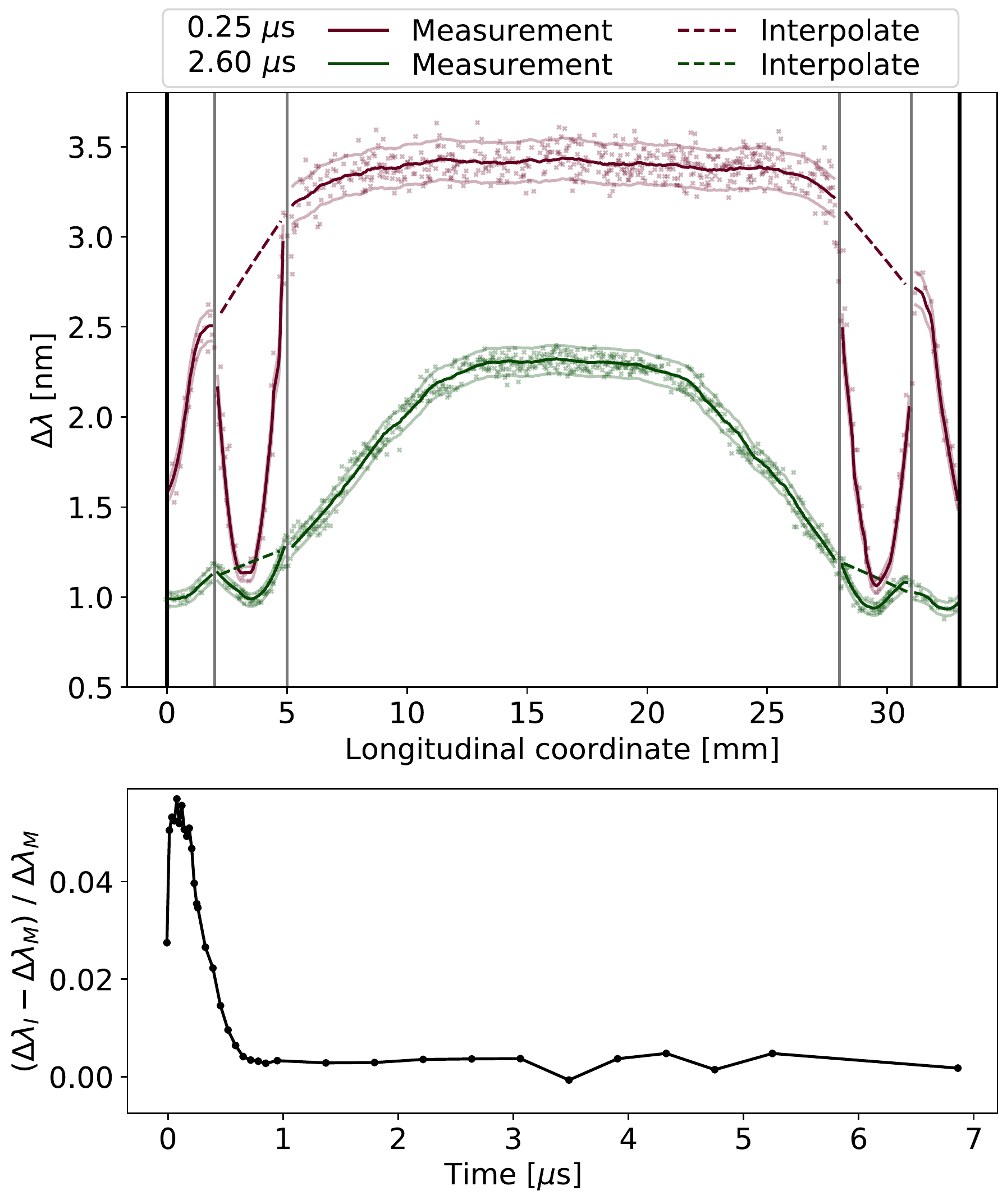}
	\caption{Red and green points show the longitudinally resolved SLB measurements $\Delta\lambda$ for 0.25 and 2.6~$\mu$s after the initiation of the discharge respectively. Each measurement was the average over 10 individual measurements at the longitudinal position $l$. The spatially smoothed (using a Savitzky-Golay routine) average is shown with a bright solid curve and the upper and lower error bounds (average error on the mean) are indicated by the weaker curves. The gas inlet regions are bounded by gray lines. A straight interpolation of the gas inlet regions is shown with dashed lines. The difference between the longitudinal average of the smoothed curves with ($\Delta\lambda_I$) and without ($\Delta\lambda_M$) interpolation in the gas inlet regions, as a function of time is shown in the lower part. ** change y-axis of lower part!**}
	\label{fig:LongFWHM}
\end{figure}

The upper part of Fig.~\ref{fig:LongFWHM} shows the longitudinally resolved SLB measurement of $\Delta\lambda$ in the 1.0~mm diameter capillary with an argon-hydrogen gas mixture of 95\% and 5\% respectively and a pressure of 40~mbar in the buffer volume, for two different times after the initiation of the discharge. The solid curves show the smoothed data using a Savitzky-Golay routine~\cite{savitzky1964smoothing} with a second order polynomial fitting. Due to the strong discontinuities around the gas-inlet regions, the Savitzky-Golay smoothing was performed in five different areas independently. The Savitzky-Golay window sizes were: 0.5~mm for 0.0~$<$~$l$~$<$~2.0 and 31.0~$<$~$l$~$<$~33.0 (the channel extremities), 1.0~mm for 2.0~$<$~$l$~$<$~5.0 and 28.0~$<$~$l$~$<$~31.0 (the gas inlet regions) and 3.0~mm for 5.0~$<$~$l$~$<$~28.0 (central part of the channel) where $l$ is the longitudinal position in mm. The effect of the focal plane shift in the gas inlet regions, i.e. the sampling of a different radial density, is visible at early times. However, this effect could not be separated from other dynamic effects such as the expulsion of plasma into the gas inlets and reduced current density in the larger volume of the gas inlet region. Therefore the true on-axis density profile in the gas inlet region is likely to lie between the measured and interpolated curves shown in Fig.~\ref{fig:LongFWHM}. In an argon plasma it was shown in simulation~\cite{sakai2011properties} that after around 300~$\mu$s the radial density profile in a capillary discharge source becomes approximately flat over about 80~\% of the radius of the capillary (from the center, outwards) and rises steeply by a maximum factor of around 1.5 at the walls. The lower part of Fig.~\ref{fig:LongFWHM} shows the absolute difference between the the longitudinal average SLB measurements given the different treatment of the gas inlet regions. The value $\Delta\lambda_M$ is the longitudinal average under the measured Savitzky-Golay smoothed curve. The value of $\Delta\lambda_I$ describes the same quantity but where the gas inlet regions are replaced by the interpolated (dashed) curve. This shows that the dynamic effects in the gas inlet regions are important in the first 500~$\mu$s after the discharge onset, but rapidly decreases thereafter. Due to the focal plane shift in the gas inlet regions, this indicates that the radial profile becomes much flatter after 500~$\mu$s for this capillary discharge. The value of $(\Delta\lambda_I-\Delta\lambda_M)/\Delta\lambda_M$ is incorporated into the systematic errors when calculating the longitudinal average of the SLB measurements as a function of time (see Sec.~\ref{sec:DiagComp}).

The methodology developed by Gigosos and Carde{\~n}oso (GC)~\cite{gigosos1996new, gigosos2003computer} was used to calculate the electron density. Their work combines experimental measurements bench-marked to an extensive set of computer-simulated SLB profiles for different hydrogen emission lines, including $H_{\alpha}$. In their work, GC calculate the simulated FWHM of SLB emission profiles $\Delta\lambda$ over a range of plasma densities $n_e$, plasma temperature $T_e$ and relative reduced mass $\mu_r$. The relative reduced mass is defined as
\begin{equation}
	\mu_r = \frac{T_e}{T_h}\mu
\end{equation}
where $T_e$ and $T_h$ are respectively the temperature of the electrons and the heavy ionic and atomic background. The reduced mass $\mu$ is defined as
\begin{equation}
	\mu = \frac{m_{emit} \cdot m_{pert}}{m_{emit} + m_{pert}}
\end{equation}
where $m_{emit}$ is the mass of the emitter particle, i.e. the atom from which the spectroscopic emission is observed, and $m_{pert}$ is the mass of the perturbing particle, i.e. the ionic/atomic background of the plasma. The complete digital data set for the $H_{\alpha}$ emission line used in references~\cite{gigosos1996new, gigosos2003computer} was obtained~\cite{GC_data} and a linear 2D interpolation method~\cite{amidror2002scattered,MATLAB:2018} was performed in log-space to connect $n_e$, $\Delta\lambda$ and $T_e$, for a given $\mu_r$. If three of these four quantities are known, the interpolation can be used to obtain the fourth quantity. For example, if $n_e$ is known from TCI measurements, $\Delta\lambda$ from SLB and $\mu_r$ simply from knowledge of the gas species, then $T_e$ can be interpolated. As the TCI diagnostic in this work yields the average on-axis plasma density and the SLB diagnostic reveals the spatially resolved broadening along that axis length $L$, the spatially averaged broadening component was computed as
\begin{equation}
	\Delta\lambda_I = 1/L\int{\Delta\lambda} dl
\end{equation}
where $\Delta\lambda$ are the individual spatially resolved SLB measurements at position $l$ in the cell (see Fig.~\ref{fig:LongFWHM}).

\section{Temperature measurement and diagnostic comparison}
\label{sec:DiagComp}

\begin{figure}[tp]
\centering
\subfigure[]{
	\includegraphics[width=1\columnwidth]{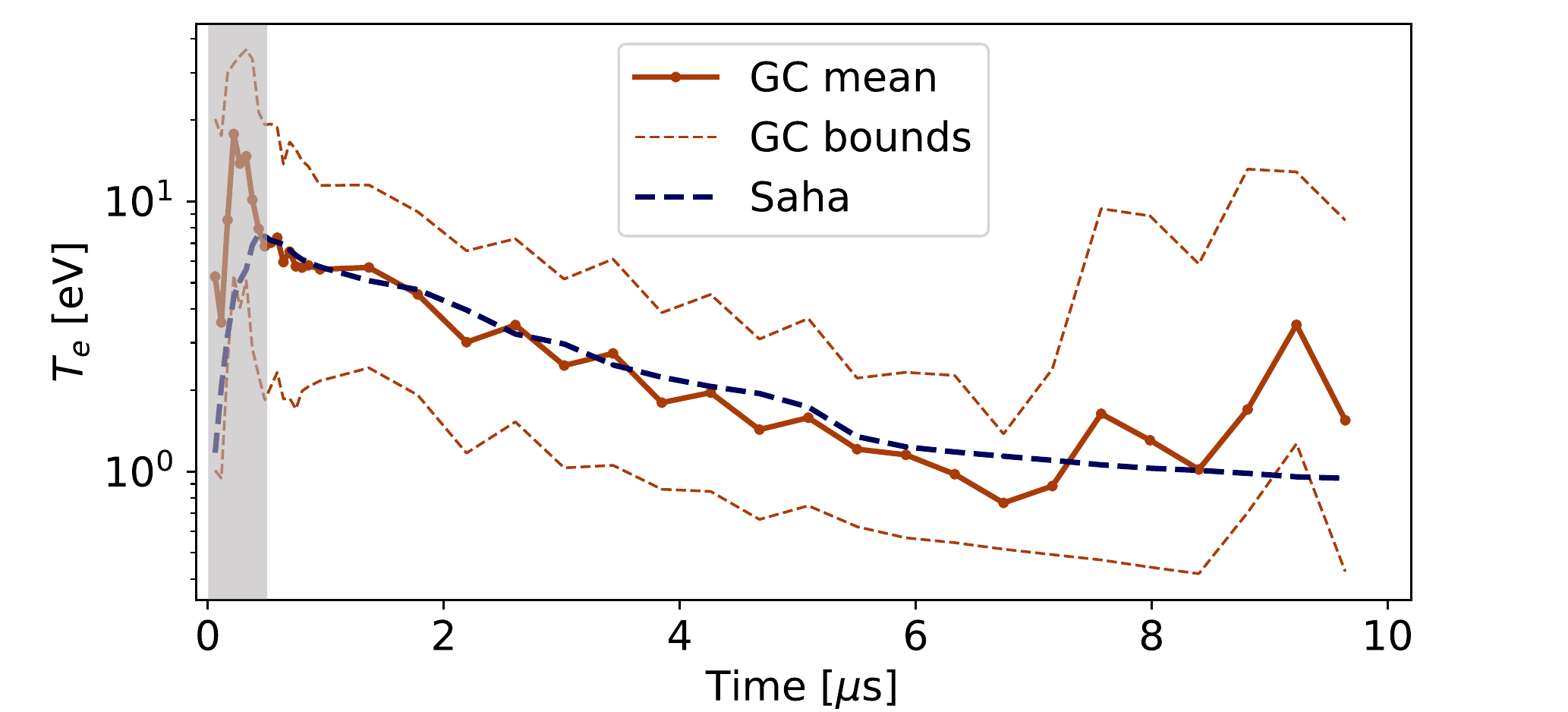}
	}
\subfigure[]{
	\includegraphics[width=1\columnwidth]{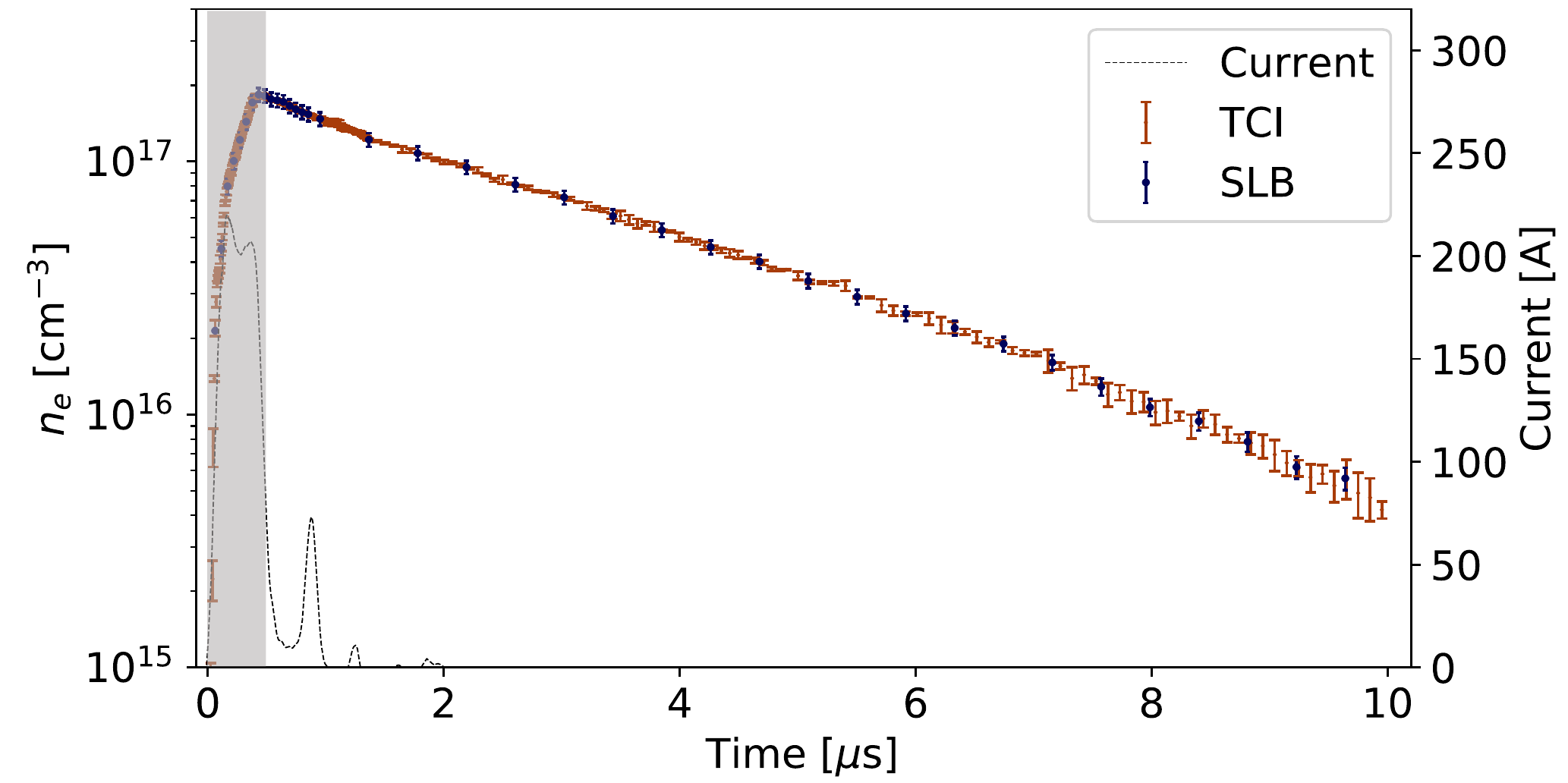}
	}
	\caption{(a) Temperature measured via the GC interpolation (orange joined points) and calculated using the Saha equilibrium theory~\cite{zaghloul2000simple} (blue dashed curve). The orange dashed upper and lower limits represent the range of possible temperatures from the GC interpolation, given the errors of the individual measurements of $n_e$ and $\Delta\lambda_I$. The grayed region represents the time for which the assumption of LTE is not likely to be valid. (b) The plasma density measurements from the SLB (blue points) and TCI (orange points). The black dashed curve shows the current profile.}
	\label{fig:Comp_result}
\end{figure}

The average on-axis temperature was computed by interpolating the GC data set, as described above. The known quantities from measurement were the average on-axis plasma density $n_e$ (TCI measurements), the average on-axis line broadening $\Delta\lambda_I$ (SLB measurements) and $\mu =$~1 for argon gas. Local thermodynamic equilibrium was assumed, hence $\mu_r =$~1. As shown by Sakai and co-authors~\cite{sakai2011properties}, the assumption of LTE conditions 0.5~$\mu$s after the initiation of the discharge is reasonable. Fig.~\ref{fig:Comp_result}~(a) shows the resulting temperature $T_e$, for the argon-hydrogen mix of 95\% and 5\% respectively at a pressure of 40~mbar measured in the buffer volume. 

The electron temperature can also be approximated from the TCI electron density measurements via the Saha ionization equations~\cite{zaghloul2000simple}, given that the atomic density $n_a$ is known. However, only the backing pressure in the buffer volume was experimentally known and not the pressure (and thus atomic density) inside the cell. To approximate $n_a$ we make the assumption that the maximum of the indirectly measured temperature (see Fig.\ref{fig:Comp_result}(a)) corresponds to the steady-state temperature for the peak current. The steady-state MHD (magnetohydrodynamic) formulations of Bobrova, Esaulov and co-workers~\cite{bobrova2001simulations, esaulov2001mhd} were solved to yield $n_a=\num{3.2e16}$~cm$^{-3}$ for the experimental peak current of 220~A and the indirectly measured peak temperature of $7.6$~eV. This value is reasonable and compatible with the TCI measurement under the assumption of LTE, considering multiple ionization of Argon (supported by the emission spectra as shown in~Fig.\ref{fig:ArSpec1}). The blue dashed line in Fig.\ref{fig:Comp_result}(a) shows the characteristic temperature evolution given by solving the Saha ionization equations and is in good agreement with the indirectly measured temperature, where LTE is assumed to be valid.

To create a more complete picture of the plasma density characterization and to use the SLB doagnostic effectively, the average on-axis plasma density for the SLB was calibrated by using the measured $\Delta\lambda_I$, the indirectly measured on-axis temperature $T_e$ and $\mu_r =$~1 (assuming LTE). The temperature-dependent measurements of $n_e$ from the SLB (orange points) are shown in Fig.~\ref{fig:Comp_result}~(b) along with those from the temperature-independent TCI (blue points). The plasma density resolution of the SLB diagnostic ($\num{5e15}$~cm$^{-3}$) can be seen in Fig.~\ref{fig:Comp_result}~(b) between 9 and 10~$\mu$s. As the temperature shown in Fig.~\ref{fig:Comp_result}~(a) agrees well with Saha                              theory and shows reasonable characteristic behavior, the combination of the two diagnostics can therefore be used with confidence to obtain accurate, spatially resolved, absolute plasma density measurements from the SLB diagnostic.

\section{Evolution of the longitudinal plasma density profile}
\label{sec:longEvolv}

\begin{figure}[!htp]
\centering
\subfigure[]{
	\includegraphics[width=1\columnwidth]{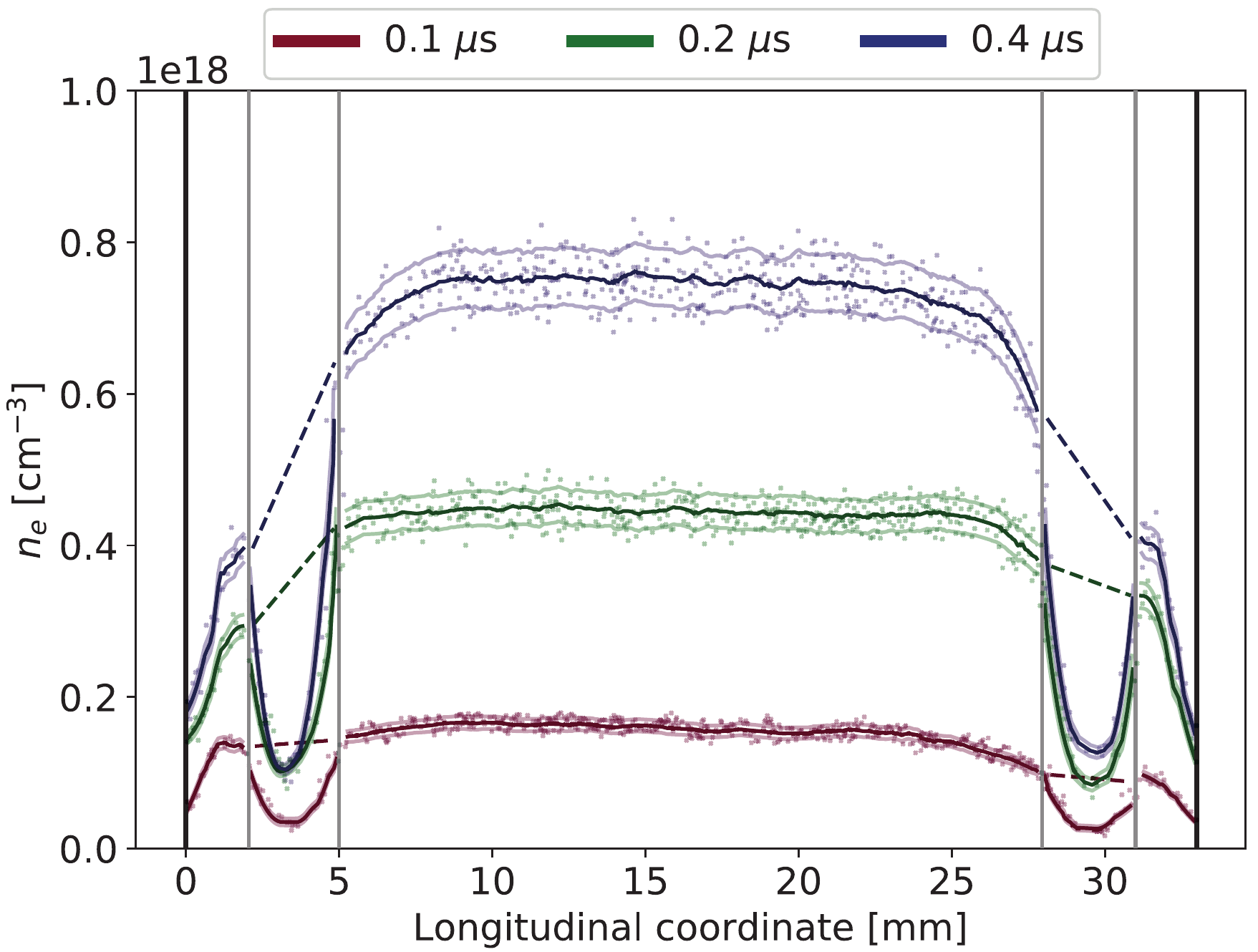}
	}
\subfigure[]{
	\includegraphics[width=1\columnwidth]{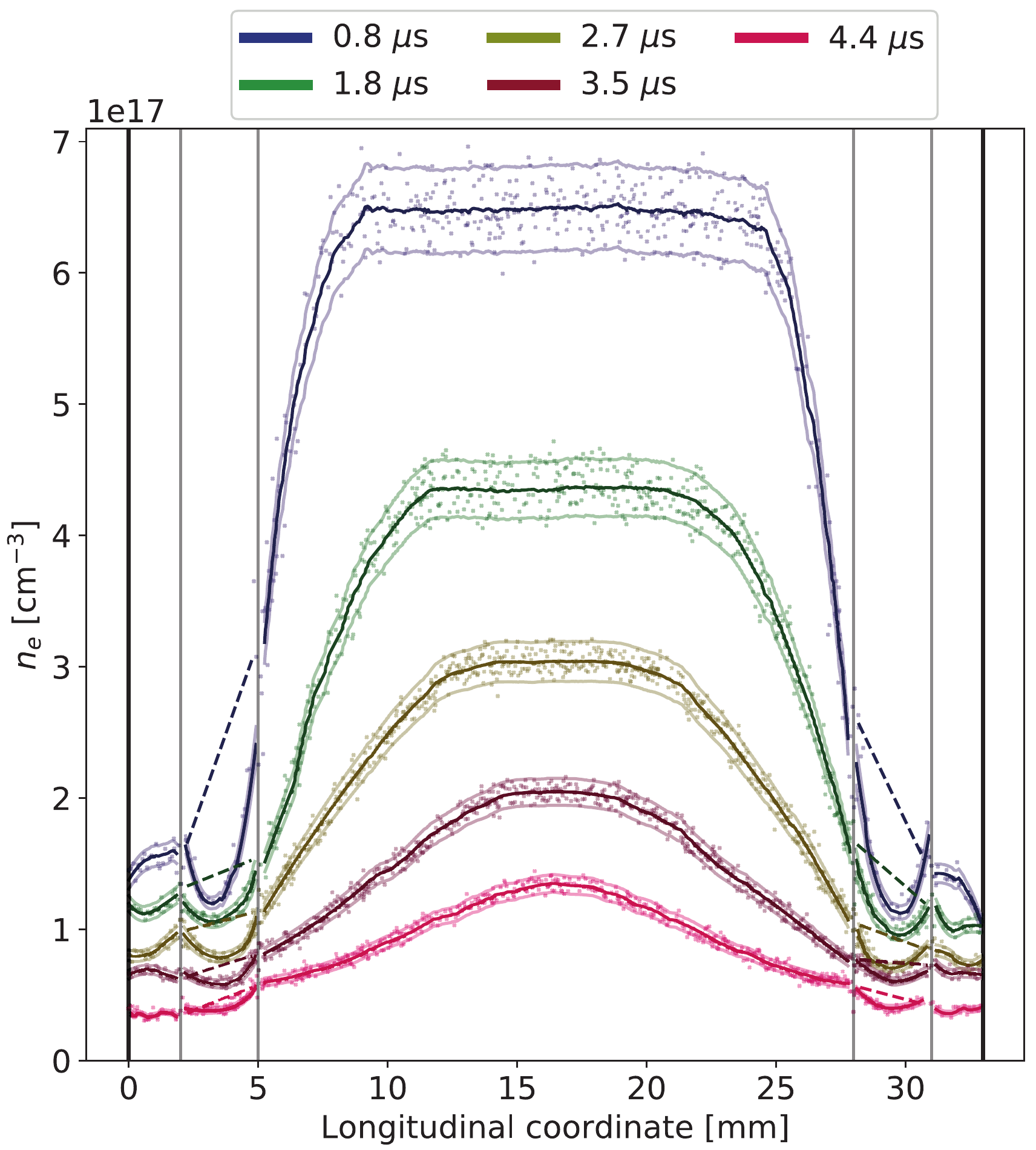}
	}
\caption{Longitudinal plasma density profiles at different times during (a) and after (b) the current pulse. The individual points show the average over 10 measurements, the Savitzky-Golay smoothing (described in Sec.~\ref{sec:Spec_diag}) is shown with a bright solid line and the upper and lower error bounds are indicated by the weaker lines above and below (95\% confidence interval of the error on the mean). The gas inlet regions are bounded by vertical gray lines. The dashed lines show a simple interpolation between the regions either side of the gas inlet.}
\label{fig:Long1_2_3}
\end{figure}

The longitudinal plasma density profile in the center of a 1.0~mm diameter capillary with the same gas mixture and pressure as in Sec.~\ref{sec:DiagComp}, was measured and using the methods described in Sec.~\ref{sec:Spec_diag} and~Sec.~\ref{sec:DiagComp}. In Fig.~\ref{fig:Long1_2_3}(a) the longitudinal profile is shown whilst the current is ramping up (red points), at the beginning of the flat-top of the current pulse (green points), and at the end of the current pulse flat-top (blue points). While the current pulse is present, the plasma density increases rapidly and with a uniform flat-top density profile in the central region, due to the initial uniform neutral gas density. 

Figure~\ref{fig:Long1_2_3}~(b) shows the further evolution of the density profile after the current pulse is switched off. The decrease in size of the central flat-top region of the profile due to expulsion of the plasma outwards from the open ends of the capillary and into the gas inlet regions is clearly shown as a function of time. The more Gaussian-like profiles shown after 2.5~$\mu$s in Fig.~\ref{fig:Long1_2_3}~(b) differ significantly from the often-desired flat-top profile shown in Fig.~\ref{fig:Long1_2_3}~(a). 

As mentioned previously, spatial (and density) resolution is reduced due to the noise present in the measurement of $\Delta\lambda$ and hence $n_e$. The dominant cause of this noise is due to jitter imposed on the iCCD camera trigger signal, resulting from the electromagnetic interference from the discharge current pulse. Small variations in the current pulse result in spurious variations in the trigger signal, hence the opening window of the intensifier in the iCCD is temporally moved from the intended position. 

The plasma density measured in the gas inlet regions is likely to be a complex combination of focal plane shift (thus radial density variation) described in Sec.~\ref{sec:Spec_diag}, reduced discharge current density due to the larger capillary volume at that location and the expulsion of plasma into the gas inlets. These effects were not disentangled in this work. However, as the plasma density measurements in these regions are lower than the central part of the capillary at early times (see Fig.~\ref{fig:Long1_2_3}~(a)), the dominant effects are likely to be reduced current density and plasma expulsion in to the gas inlets. If these effects were not dominant then the effect of the focal plane shift would result in a higher plasma density measurement in these regions, due to the more strongly pronounced parabolic radial density profile in argon plasma in the first few 100's of ns~\cite{sakai2011properties}. The difference between the gas inlet regions and the rest of the capillary becomes smaller with time (see Fig.~\ref{fig:Long1_2_3}~(b)) as the discharge current ceases and the plasma is no longer non-uniformly heated or expelled into the gas inlet regions. The dashed part of the curves in Fig.~\ref{fig:Long1_2_3} represent a reasonable maximum on-axis density in the gas inlet regions and therefore the maximum reasonable error. Ongoing studies are being performed to measure the radial density profile and understand in detail the effect the gas inlets have on the on-axis density profile.

\section{Conclusions}
\label{sec:Discuss}

Two complementary plasma diagnostics were used to characterize the plasma density and temperature in discharge capillaries: a transversely-mounted spectrometer for recording broadened plasma emission spectra with a lower density resolution of $\num{5e15}$~cm$^{-3}$ and a longitudinal common-path two-color laser interferometer with a sensitivity of $\num{2.0e15}$~cm$^{-2}$. The TCI was used together with the SLB diagnostic to indirectly measure the average on-axis temperature evolution of the plasma. The temperature exhibits reasonable characteristic behavior which agrees with theory. 

The sensitivity was improved in this work compared to previous studies of plasma density in discharge capillaries. Previous studies were made using spectral broadening techniques in the plasma density range $\num{1e18}$ to $\num{1e19}$~cm$^{-3}$~~\cite{ashkenazy1991spectroscopic, edison1993characterization, curcio2019modeling}, with some reporting lower densities of $\num{1e17}$~~\cite{ashkenazy1991spectroscopic}. Most such studies were also made with pure hydrogen plasmas and collected spontaneous light emission along the longitudinal capillary axis, as opposed to perpendicular to it. Together with higher plasma densities ($\num{1e18}$ to $\num{1e19}$~cm$^{-3}$), this increases the likelihood of self-absorption becoming a problem. This may explain why comparison studies carried out between spectrometry and interferometry in discharge capillaries~\cite{jang2011density} have exhibited diagnostic disagreement. Typically in previous studies of discharge capillaries utilizing spectral line broadening, the temperature is simply estimated as roughly 1~-~3~eV. However, in the work presented in this paper the temperature is extracted using the combination of temperature-dependent and -independent plasma diagnostics. This removes the necessity of ad hoc guesswork and allows for the temperature to evolve naturally in time, improving the accuracy of the method.

The importance of measuring the evolution of the longitudinal profile in a discharge capillary plasma source was highlighted. Figure~\ref{fig:Long1_2_3} shows clearly that the longitudinal profile evolves from a so-called flat-top to a more Gaussian-like shape, during which the extremities of the plasma distribution are expelled from the ends of the plasma source and into the gas inlet regions. This kind of measurement is of the utmost importance when considering the beam dynamics in a plasma wakefield accelerator due to the evolution of the density and hence wakefield strength that the beam will experience in its transit through the capillary. With precise knowledge of the longitudinal profile and its temporal evolution, suitable plasma wakefield accelerator parameter choices can be made. For example at FLASHForward, the combination of initial neutral-gas pressure, discharge current and voltage, and the wakefield drive-beam arrival time with respect to a discharge. Tuning these parameters presents the possibility to obtain a specific profile at a given plasma density with suitable entrance and exit ramps to preserve emittance and minimize beam hosing. Such precise characterization will enable better control over self-injection methods and dephasing for laser- and beam-driven plasma wakefield accelerators.

\section{Acknowledgements}

The authors would like to acknowledge funding by the Helmholtz Matter and Technologies Accelerator Research and Development Program and the Helmholtz IuVF ZT-0009 grant. We would also like to thank Prof. Gigosos for his invaluable help in compiling the data needed to complete the analysis.

\section{Data Availability}

The data that support the findings of this study are available from the corresponding author upon reasonable request.

\bibliography{refs}

\end{document}